# Econometric Model Using Arbitrage Pricing Theory and Quantile Regression to Estimate the Risk Factors Driving Crude Oil Returns

Sarit Maitra[a], Vivek Mishra[b], Sukanya Kundu[c], Manav Chopra[d]

[a, c, d] *Business School, Alliance University, Bengaluru, 562106, India*
[b] *School of Applied Mathematics, Bengaluru, 562106, India*
*Corresponding author: sarit.maitra@gmail.com*

*Abstract*— **This work adopts a novel approach to determine the risk and return of crude oil stocks by employing Arbitrage Pricing Theory (APT) and Quantile Regression (QR). The APT identifies the underlying risk factors likely to impact crude oil returns. Subsequently, QR estimates the relationship between the factors and the returns across different quantiles of the distribution. The West Texas Intermediate (WTI) crude oil price is used in this study as a benchmark for crude oil prices. WTI's price fluctuations can have a significant impact on the performance of crude oil stocks and, subsequently, the global economy. To determine the proposed model's stability, various statistical measures are used in this study. The results show that changes in WTI returns can have varying effects depending on market conditions and levels of volatility. The study highlights the impact of structural discontinuities on returns, which can be caused by changes in the global economy and the demand for crude oil. The inclusion of pandemic, geopolitical, and inflation-related explanatory variables add uniqueness to the study as it considers current global events that can affect crude oil returns. Findings show that the key factors that pose major risks to returns are industrial production, inflation, the global price of energy, the shape of the yield curve, and global economic policy uncertainty. This implies that while making investing decisions in WTI futures, investors should pay particular attention to these elements.**

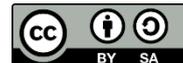

## 1. INTRODUCTION

International crude oil prices have significantly fluctuated in recent years, largely due to factors such as global economic conditions, technological advancements, political instability, and natural disasters. Despite hundreds of oil production locations, only a few crude oil benchmarks are used for oil pricing: WTI[1] and Brent. The prices of these benchmarks have played a significant role in the variations in prices. This study aims to provide a new way of analyzing the risk and return of crude oil stocks in an uncertain market by combining market fundamentals and economic factors. The literature on multifactor analysis for oil returns is limited, as most research has focused on univariate correlations between oil prices and a single factor. The study uses a multi-factor QR model combined with APT to provide a more comprehensive understanding of oil prices.

The use of APT and QR to calculate the risk and return on crude oil stocks is an intriguing and profitable technique. APT is a multifactor model, which means it considers a variety of risk factors that affect any asset's returns. The fundamental tenet of APT is that an asset's expected return is a linear function of how exposed it is to different risk factors plus a particular risk premium (idiosyncratic risk) that is unique to that asset. According to the hypothesis, investors will constantly try to take advantage of arbitrage possibilities to correct market inaccuracies and create equilibrium situations. The existence of transient and short-lived arbitrage opportunities in financial markets is not denied by even the Efficient Market Hypothesis. The occurrence of arbitrage

---

[1] WTI is a blend of several domestic crude streams in the United States, with its main trading location in Cushing, Oklahoma. Brent crude oil encompasses four crude streams pumped in the North Sea.

opportunities is a critical mechanism that contributes to market efficiency.

The APT strategy aims to identify risk factors influencing WTI returns and use them to explain predicted returns. By considering multiple factors simultaneously, it aims to uncover hidden relationships and interactions. QR is a distribution analysis technique used to estimate the conditional distribution of data at different quantiles. While the individual methods are not new, their combination and application to crude oil stocks in conjunction with each other is a novel approach. Combining these methods with crude oil stocks is a novel approach. Descriptive statistics show relevant distributions are leptokurtic, justifying quantile regression to detect herding bias in tails. Herding bias refers to investors imitating others' actions without independent analysis.

The two main contributions to this paper are:
- implementation of APT to identify the underlying risk factors driving ROC,
- application of QR to estimate the effect of these risk factors on different segments of the distribution of WTI returns.

QR provides a deeper insight into the nature of the relationship between the factors and the returns. In a recent study, Zhao et al. (2023), emphasized the importance of investor sentiments, taking a clue from behavioral finance (Aloui et al. 2020). As a result, a multiple-factor study considering the current state of the oil industry is a crucial research agenda item. The results show distinctive and original aspects of the impact, such as industrial production (PROD), INFLATION, global price of energy (GPE), and global economic policy uncertainty (GEPU), and they have several implications for investors and decision-makers to reduce investment risks

## 2. PREVIOUS WORK

We find the studies on the relationship between multiple-factors and the ROC (return on crude oil) are limited, which indicates a potential gap in the literature. However, there is a growing trend of academic research on macroeconomic factors impacting ROC (e.g., McMillan et al. 2021; Nayar 2020; Hamdi et al. 2019; Salisu et al. 2019, etc.). Additionally, other aspects have been studied by researchers, such as the relationship between the pandemic and crude oil (Liu et al. 2020; Prabheesh et al. 2020, etc.) and geopolitical unpredictability (Wei et al. 2019; Alqahtani et al. 2020). In recent academic works, researchers have studied numerous risk factors to evaluate WTI returns such as:
- Macroeconomic indicators: GDP, inflation, interest rates, and consumer sentiment are some of the macroeconomic variables that have been found to be significant predictors of WTI returns (McMillan et al. 2021; Mokni 2020; Nayar 2020).
- Geopolitical events: Political events like wars, terrorist attacks, and geopolitical conflicts significantly influence oil prices (McMillan et al. 2021; Shahzad et al. 2021; Mahmoudi & Ghaneei 2022; Wei et al. 2019; Alqahtani et al. 2020).
- Financial market indicators: It has also been discovered that factors including stock market indices, volatility, and credit spreads can accurately predict oil prices (McMillan et al. 2021; Shahzad et al. 2021).
- Energy policies: Government policies related to energy production, consumption, and conservation can also impact oil prices (McMillan et al. 2021).

These studies provide insights into the various factors that have been considered to estimate WTI returns during the last decade, such as, economic indicators, geopolitical events, exchange rates, financial market indicators, and energy policies. Some of the most researched factors have been:
- US Treasury Spread: Several research have discovered a considerable effect of the US Treasury yield spread on crude oil prices (e.g., Dai and Kang, 2021; Ferrer et al., 2018 etc.). While other research (such as Guo et al., 2021) suggest that a larger spread causes higher oil prices, Wang et al. (2023) discovered that the impact is time varying.
- Global economic policy uncertainty: There has been substantial progress in our understanding of how uncertainty in oil prices, economic policy, and overall economic activity are related. According to research by Shahzad et al. 2019, Herrera et al. 2019, Adekoya et al. 2022, etc., there is a correlation between the volatility of the oil price and the unpredictability of global economic policy.
- Inflation: Inflation can impact oil prices by affecting the demand for oil as well as the cost of production (Husaini & Lean 2021; Köse & Ünal 2021, etc.).
- Industrial production: Changes in industrial production can affect the demand for oil, as industrial processes often rely on oil as an input (Singhal et al., 2019; Wei et al., 2019; Herrera et al., 2019).
- Currency fluctuation with the euro: The study by Malik and Umar (2019) finds that changes in exchange rate volatility are not explained by changes in oil prices. However, they have discovered a strong link between the volatility of currency rates. Salisu et al. (2022) assert that fluctuations in the value of the US dollar are directly influenced by changes in the price of oil.
- Narrow money supply: As changes in the money supply influence total economic activity, they can also influence the demand for oil. According to Lee et al. (2019), there is evidence of a correlation between the volatility of the oil price and the unpredictability of economic policy.
- Unemployment rate: Empirical results indicate a dynamic causal link between unemployment and ROC (Wang et al. 2022). Chan & Dong (2022) found that an unanticipated rise in oil price volatility causes the jobless rate to persistently rise.
- VIX: Chen (2022) reveals that investment horizons impact EPU, VIX, and GPR's influence on oil stock movement. VIX is the most significant uncertainty measure in developed markets, while Brazil, India, GPR, and EPU are vulnerable in emerging markets.

WTI returns were also significantly impacted by the COVID-19 epidemic, with prices collapsing as demand fell as a result of travel bans and lockdowns. Oil prices and COVID-19 instances have a negative link, claim Chua & Yang (2021). A growing body of research (e.g., Salisu et al. 2019, Pan et al. 2017, Le & Chang 2013) points to a non-linear connection between oil prices and economies despite the studies' primary emphasis being on linear models. Nonlinearities, in accordance with Beckmann & Czudaj (2013), may be brought on by major external oil price shocks, discrete regime changes, or the inherently nonlinear structure of the data generation method (Alqaralleh 2020). In the literature, there is no agreement on the best effective approaches for doing multi-factor analysis and diagnostic testing for WTI excess returns.

Despite numerous research on the relationship between oil prices and macroeconomic factors, the literature on the risk and return characteristics of crude oil assets in respect to these variables is still lacking. While earlier research has focused on the relationship between oil prices and macroeconomic indicators (e.g., McMillan et al. 2021; Nayar 2020; Hamdi et al. 2019, etc.), it has not adequately investigated the implications of these findings on the risk and return characteristics of crude oil stocks.

Hence, the above review provides a clear argument for more research into the relationship between several parameters and crude oil returns. While there has been some research on individual factors such as macroeconomic indicators, geopolitical events, financial market indicators, energy policies, and the impact of COVID-19 on oil prices, there is a gap in the literature regarding the comprehensive analysis of all these factors in a single crude oil return model.

Table 1 displays a bibliometric report which analyzes the citation patterns and impact of scholarly articles in the field. The citation analysis indicates the influence and popularity of the studies in the field.

TABLE 1
BIBLIOMETRIC REPORT

| Sl. No | Authors | Journals | citations |
|---|---|---|---|
| 1 | Ferrer et al. (2018) | Elsevier (Energy Economics), https://doi.org/10.1016/j.eneco.2018.09.022 | 350 |
| 2 | Singhal et al. (2019) | Elsevier (Resources Policy) https://doi.org/10.1016/j.resourpol.2019.01.004 | 232 |
| 3 | Prabheesh et al. (2020) | Energy Research Letters, 1(2). https://doi.org/10.46557/001c.13745 | 227 |
| 3 | Pan et al. (2017) | Elsevier (Journal of Empirical Finance) https://doi.org/10.1016/j.jempfin.2017.06.005 | 161 |
| 4 | Beckmann & Czudaj (2013) | Elsevier (International Review of Economics & Finance) https://doi.org/10.1016/j.iref.2012.12.002 | 156 |
| 5 | Le & Chang (2013) | Elsevier (Energy Economics) https://doi.org/10.1016/j.eneco.2012.12.002 | 153 |
| 6 | Wei et al. (2019) | Elsevier (Finance Research Letters) https://doi.org/10.1016/j.frl.2019.03.028 | 145 |
| 7 | Herrera et al. (2019) | Elsevier (Energy Policy) https://doi.org/10.1016/j.enpol.2019.02.011 | 131 |
| 8 | Malik and Umar's (2019) | Elsevier (Energy Economics) https://doi.org/10.1016/j.eneco.2019.104501 | 126 |
| 9 | Hamdi et al. (2019) | Elsevier (Energy Economics) https://doi.org/10.1016/j.eneco.2018.12.021 | 123 |
| 10 | Salisu et al. (2019) | Elsevier (Economic Modelling) https://doi.org/10.1016/j.econmod.2018.07.029 | 103 |
| 11 | Shahzad et al. (2021) | Elsevier (International Review of Financial Analysis) https://doi.org/10.1016/j.irfa.2021.101754 | 86 |
| 12 | Alqahtani et al. (2020) | Elsevier (Economic Analysis and Policy) https://doi.org/10.1016/j.eap.2020.09.017 | 53 |
| 13 | Mokni (2020) | Elsevier (Energy) https://doi.org/10.1016/j.energy.2020.118639 | 31 |
| 14 | Alqaralleh (2020) | Taylor & Francis (Journal of Applied Economics) https://doi.org/10.1080/15140326.2019.1706828 | 30 |
| 15 | Köse & Ünal (2021) | Elsevier (Energy) https://doi.org/10.1016/j.energy.2021.120392 | 29 |
| 16 | Husaini & Lean (2021) | Elsevier (Resources Policy) https://doi.org/10.1016/j.resourpol.2021.102175 | 24 |
| 17 | Dai and Kang (2021) | Elsevier (Energy Economics), https://doi.org/10.1016/j.eneco.2021.105205 | 21 |
| 18 | Wang et al. (2022) | Elsevier (Energy) https://doi.org/10.1016/j.energy.2022.124107 | 16 |
| 19 | Mahmoudi & Ghanei (2022) | Emerald Publishing Limited (Studies in Economics and Finance) https://doi.org/10.1108/SEF-09-2021-0352 | 15 |
| 20 | McMillan et al. (2021) | Elsevier (Energy Economics) https://doi.org/10.1016/j.eneco.2021.105102 | 12 |
| 21 | Chan & Dong (2022) | Elsevier (Economic Modelling) https://doi.org/10.1016/j.econmod.2022.105935 | 11 |
| 22 | Salisu et al. (2022) | Elsevier (Energy Economics), https://doi.org/10.1016/j.eneco.2022.105960 | 9 |
| 23 | Wang et al. (2023) | Science Direct, Borsa Istanbul Review https://doi.org/10.1016/j.bir.2022.12.003 | 9 |
| 24 | Nayar (2020) | Emerald (International Journal of Energy Sector Management https://doi.org/10.1108/IJESM-08-2018-0006 | 6 |

| 25 | Adekoya et al. (2022) | Elsevier (Resources Policy) https://doi.org/10.1016/j.resourpol.2022.103004 | 2 |
| 26 | Chen (2022) | Taylor & Francis (Applied Economics) https://doi.org/10.1080/00036846.2022.2140115 | 1 |

The literature review presents a compelling argument for the significance of conducting research on various factors and their influence on the risk and returns associated with crude oil. It underscores the imperative need for additional scholarly inquiry to deepen our comprehension of the intricate interaction between these factors and crude oil dynamics. The review identifies critical research areas that warrant investigation, thereby paving the way for future academic endeavors to fill the existing gaps in knowledge and contribute to the advancement of this field of study

### 3. MODEL AND ECONOMETRIC APPROACH

The APT model provides a framework for understanding asset pricing based on the systematic risk factors that influence asset returns. The equilibrium asset pricing equation according to the APT model is:

$$E(R_i) = Rf + \beta_{1i} * f_1 + \beta_{2i} * f_2 + \cdots + \beta_{ki} * f_k \quad (1)$$

Where, $E(R_i)$ is the expected return of asset $i$, $Rf$ is the risk-free rate of return, $\beta_{1i}, \beta_{2i}, \ldots, \beta_{ki}$ are the sensitivity coefficients of WTI $i$ to the $k$ systematic risk factors $(f_1, f_2, \ldots, f_k)$, which represent different sources of risk in the economy.

The βs are estimated by using linear regression. This is calculated using QR with an estimate of the conditional median (0.5 quantile), and the model's adequacy was checked using various regression diagnostic tests. The coefficients of the regression represent the sensitivities of the asset to each factor, while the intercept term represents the risk-free rate of return.

*A. Data source and variables*

The study used five years of monthly data from the FRED Economic Database to estimate betas, using yield data for crude oil, specifically the WTI crude oil spot price, to provide insights. The S&P 500 index, represented by the SPY, was chosen as the benchmark for the US equity market and a symbol of financial stability and US economy health. Table 2 presents selected variables, macroeconomic indicators, and other relevant data sources for the research question.

TABLE 2
VARIABLE SELECTION & DATA SOURCE

| Factors | Characterization variable | Abbreviation |
|---|---|---|
| U.S. Treasury Securities at 3-Month Constant Maturity | DGS3MO index | DGS3MO |
| U.S. Treasury Securities at 5-year Constant Maturity | DGS5 index | DGS5 |
| Industrial Production: Total Index | INDPRO index | PROD |
| Consumer Price Index for All Urban Consumers | CPIAUCSL | INFLATION |
| Unemployment rate | UNRATE index | UNRATE |
| Narrow money supply | M1SL index | M1SL |
| Change in exchange rate | CCUSMA02EZM618N index | CCU |
| S&P 500 index | SP index | SP |
| CBOE Market Volatility Index | VIX index | VIX |
| Geopolitical Risk Index | GPR data | GPR2 |
| Global price of Energy index | PNRGINDEXM index | GPE |
| World Pandemic Uncertainty Index | WUPI index | WUPI |
| Global economic policy uncertainty | GEPU index | GEPU |
| International crude oil price | WTI crude oil spot price | WTI |

Note: Given S&P's reduction of the nation's credit rating in 2011, the common perception that U.S. treasury securities are devoid of credit risk may be debatable; nonetheless, that subject is outside the purview of this study.

The US Treasury bill rate is viewed as a risk-free interest rate due to its empirical relevance and theoretical basis in explaining crude oil returns. Here are some reasons why these factors have been chosen:

- **SPREAD**: The yield spread is a measure of risk and investor sentiment, reflecting market expectations for future economic conditions. It can be used to capture market sentiment and its impact on crude oil returns, as oil prices are sensitive to economic changes. The US Treasury spread is particularly useful in this context.
- **GEPU**: The study explores the correlation between oil price volatility, economic policy uncertainty, and crude oil returns, highlighting the significant impact of these factors on investment decisions, global economic growth, and geopolitical stability.
- **INFLATION**: Inflation affects the purchasing power of consumers and can impact the demand for oil. Higher inflation may increase production costs and reduce consumer spending, potentially affecting oil prices. We have explored the impact of inflation on crude oil returns by including it as a factor.
- **PROD**: Changes in industrial production can reflect overall economic activity and demand for oil. Industries heavily reliant on oil as an input may experience fluctuations in production levels, which can, in turn, influence oil prices. Understanding the relationship between industrial production and crude oil returns requires taking it into account as a component.

---
[2] GPR data obtained from https://www.matteoiacoviello.com/gpr.htm

- **CCU:** Exchange rate fluctuations, particularly with major currencies like the euro, can affect oil prices. This can impact the affordability of oil for different countries and influence demand. Including currency fluctuation as a factor allowed us to explore the relationship between exchange rates and crude oil returns.
- **M1SL:** Changes in the money supply can have an impact on overall economic activity, which, in turn, can affect the demand for oil. By considering the narrow money supply, we have examined its relationship with crude oil returns and assessed its influence on oil market dynamics.
- **UNRATE:** This reflects labor market conditions and can indicate the overall economic health. High unemployment may affect consumer spending and demand for oil. Incorporating the unemployment rate as a consideration helped in comprehending the relationship between crude oil returns.
- **VIX:** The VIX index measures market volatility and investor sentiment. High levels of market volatility can impact oil prices as they affect investor risk appetite and their investment decisions. We have investigated the VIX's impact on crude oil returns by including it as a factor in the analysis.

We can classify the variables into two categories: market fundamentals and economic indicators where, INDPROD, M1SL, CCU, SP, GPR, GPE, WUPI, and GEPU; and DGS3MO, DGS5, CPIAUCSL, UNRATE and VIX

*B. Econometric approach*

This work is divided into two stages: the first stage examines the excess return over time, and the second stage analyses the excess return's cross-section components. Our study's major presumptions are that markets are efficient, events cannot be predicted, and time is affected exogenously.

Eq. (1) can be extended to Eq. (2) to discuss the excess return on WTI.

$$R_{wti} = \alpha_1 + \beta_M R_{Mt} + \beta_1 SPREAD_t t + \beta_2 INDPRO_t \\ + \beta_3 INFLATION_t + \beta_4 UNRATE_t \\ + \beta_5 M1SL_t + \beta_6 CCUS_t + \beta_7 VIX_t \\ + \beta_8 GPR_t + \beta_9 WUPI_t + \beta_{10} GPE_t \\ + \beta_{11} GEPU_t + \varepsilon_t$$
(2)

Here,
- $R_{wti}$ = the excess ROC; the risk-free rate which is 3-month US Treasury bill rate here, was deducted from the continuously compounded returns to transform the WTI returns into excess returns,
- $R_{Mt}$ = the excess market return on the stock; thus, excess ROC is affected by the excess market return $R_{Mt}$, and the coefficient $\beta_M$. We have used SP500 as excess market return. $SP_t$ = the excess market return,
- $SPREAD_t$ = 5-year minus 3-months treasury yield curve,
- β = coefficient,
- $\varepsilon_t$ = error term.

Rest all are the important risk factors that have been identified in the literature as having an impact on WTI returns.

## 4. METHODOLOGY

According to the APT, the focus is on unanticipated changes in macroeconomic factors rather than their levels to explain stock returns. In line with this principle, several modifications are made to the variables. The study makes the naive assumption that investors' expectations for the future value of the variables will remain unchanged. Thus, the unforeseen change is the overall variation in the variable from one period to the next. Eq. (3) displays the calculation of the monthly logarithmic excess returns for WTI, where the 3-month U.S. Treasury rate is used as the risk-free rate.

$$ER_{wti\ (t)} = \ln(p_{wti\ (t)} / p_{wti\ (t-1)}) - r_f \quad (3)$$

In Eq. (3),
- $EER_{wti\ (t)}$ is the excess return of WTI at time $t$,
- $p_{wti\ (t)}$ is the price of WTI at time $t$,
- $p_{eti\ (t-1)}$ is the price of WTI at time $t - 1$, and
- $r_f$ is the 3-month U.S. Treasury rate.

The monthly excess returns are calculated by subtracting the monthly yield on a three-month US T Bill from the continuously compounded daily returns on the WTI Index. The log changes of the data are used to express the macroeconomic elements that function as predictors. The VIX and SP are both expressed in levels.

We have used Eq. (4) to calculate the daily log changes:

$$VIX_t = \ln(VIXt_t / VIX_{t-1})\ and\ SP_t = \ln(SP_t / SP_{t-1}) - r_f \quad (4)$$

where, $VIX_t$ and $SP_t$ are the daily log change at time $t$, $VIX_t$ and $SP_t$ are the values at time $t$, and $VIX_{t-1}$ and $SP_{t-1}$ are the values at time $t - 1$.

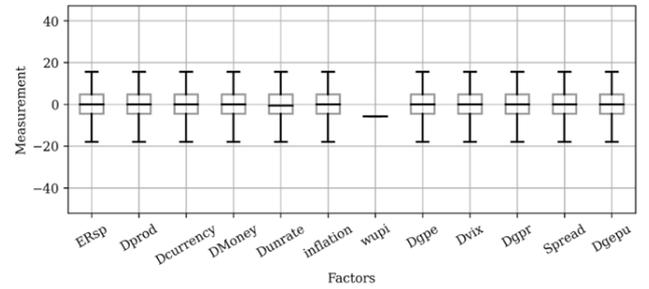

Fig. 2. Boxplot of data after Quantile Normalization

Rest all the factors are standardized using $\{\log Diff\ (X_t) - \log Diff\ (X_{t-1})\}$, where X is the respective factor and Xt-1 is the value at the previous time. Appendix presents the descriptive statistics for all the variables. The table displayed in the Appendix comprises the variance inflation factor (VIF), the Augmented Dickey-Fuller (ADF) (Dickey & Fuller 1979) unit root test empirical statistics, and the Jarque-Bera (JB) test for the normality assumption. The VIF values for the independent variables are all < 3.0, indicating that multi-collinearity is not present (Hair et al. 2017). The ADF-tests demonstrate that all series are stationary. Many of the variables show skewness and a high amount of kurtosis. Large excess kurtosis coefficients, which is leptokurtosis, are a sign that outliers are present and indicate that there have been numerous price changes in the past (either positive or negative) away from the average returns for the investment. Despite a positive mean reflecting favorable results on the average return for investors, the negative skewness (Fig. 1) shows that more negative data is concentrated on the mean value.

According to the large standard deviation values associated with various variables, the pandemic crisis of 2020–21 and the post–crisis period makes up over half of the data set. At the 5% significance level, the JB test statistics reject the null hypothesis ($H_0$) of a normal distribution for all series.

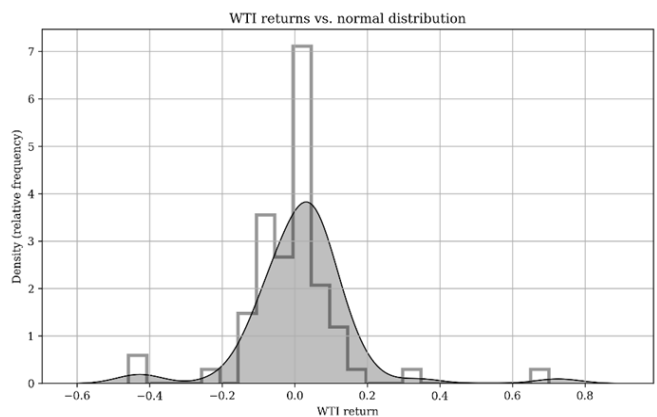

Fig. 1. Shewed target distribution

Considering the minimum values, the lowest in this range is UNRATE, with a minimum value of -131.38. GEPU is much more dispersed than other variables, with a standard deviation of 43.40; closely following this are the GPR with 30.93, UNRATE with 21.29, and MONEY with 20.32. Negative values for skewness are common (SP, CURRENCY, MONEY, UNRATE, INFLATION, GPR, and SPREAD) but are positive for the INDPRO, PANDEMIC, GPE, VIX, and GEPU. Most of these factors show excess kurtosis. To develop a new coordinate system and align it with the largest variation in the data, Principal Component Analysis (PCA) was carried out. The results are displayed in the next section.

Value at Risk (VaR) is estimated (Table 3) on simple returns, which represent the worst-case loss associated with probabilities, and CVaR is estimated by averaging the severe losses in the tail of the distribution of WTI returns.

TABLE 3
WTI VALUE AT RISK

| ` | VaR | Conditional VaR |
|---|---|---|
| 90% | -0.10 | -0.22 |
| 95% | -0.13 | -0.30 |
| 99% | -0.43 | -0.43 |

The quantile normalization procedure was used to modify the raw data to preserve the true variance that we were interested in while removing any unwanted variation induced by technological artefacts. The normalized box plot of the dependent variables is shown in Fig. 2.

The proportion of eigenvalues attributed to each component is shown in Fig. 3. This indicates the importance of each component for the analysis.

## 5. MULTIFACTOR QUANTILE ESTIMATES

Table 4 reports the regression estimation ($Qn0.5$) based on Eq. 4. The diagnostic tests were performed on the conditional median quantile, which has been treated here as the estimation results for the baseline regression.

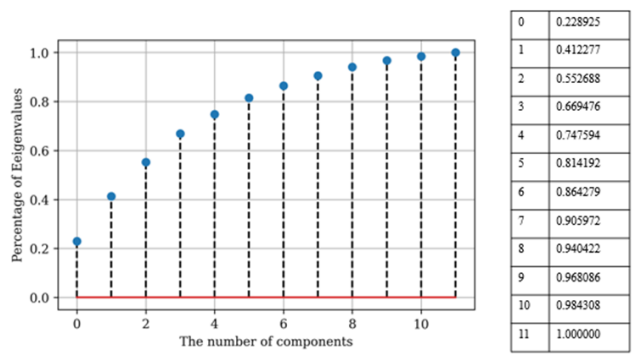

Fig. 3. Percentage of Eigenvalues Attributable to Each Component

TABLE 4
REGRESSION ESTIMATION

| Variable | Coef | Std err | $P > |t|$ |
|---|---|---|---|
| erSP | -0.17 | 0.07 | 0.01 |
| dPROD | 0.14 | 0.06 | 0.03 |
| dCURRENCY | 0.04 | 0.04 | 0.36 |
| dMONEY | -0.10 | 0.06 | 0.09 |
| dUNRATE | 0.30 | 0.04 | 0.00 |
| dINFLATION | 0.12 | 0.07 | 0.07 |
| dWUPI | -0.04 | 0.04 | 0.32 |
| dGPE | 0.34 | 0.06 | 0.00 |
| dVIX | -0.47 | 0.08 | 0.00 |
| dGPR | 0.06 | 0.05 | 0.20 |
| dSPREAD | 0.06 | 0.05 | 0.26 |
| dGEPU | 0.08 | 0.04 | 0.04 |
| Intercept | -0.01 | 0.28 | 0.96 |
| Pseudo R2 | 0.5185 | | |

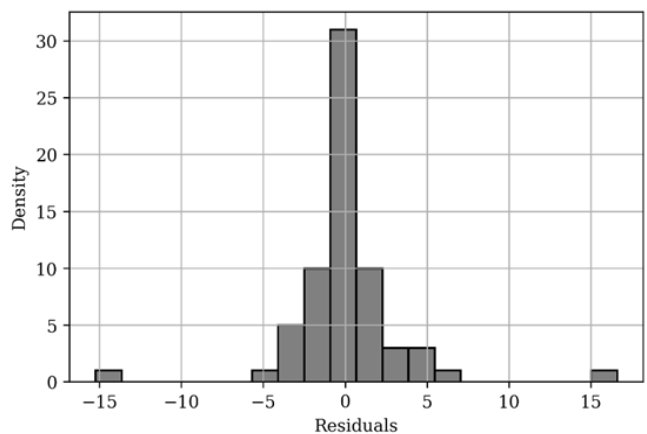

Fig. 4. Histogram of residuals

The asymmetry in the model can be seen by comparing the coefficients of various quantiles. A few parameter estimations, e.g., "$Dcurrency$", "$DMoney$", "$inflation$", "$wupi$", "$Dgpr$", and "$Spread$" variables, are not statistically distinct from zero.
$$(hypotheses = "Dcurrency = DMoney = inflation = wupi = Dgpr = Spread = 0").$$
'D' prefix added to the relevant dataset after differencing to stabilize the series.

Using an F-test, we tested $H_0$ that the parameters on these six variables are all zero. The resulting F-test statistic value is 2.98, with a p-value of 0.013 indicating the regression model better fits the data than the model with no independent variables. This result is promising because it demonstrates that the independent variables in our model improve the model's fit.

Heteroscedasticity was assessed using the Breusch-Pagan test. Table 5 reports the test results.

TABLE 5
BREUSCH-PAGAN HETEROSKEDASTICITY TEST

| Test statistic | p-value | f-value | f(p-value) |
|---|---|---|---|
| 46.71 | 0.000 | 10.6969 | 0.000 |

p-values < 0.05, indicating a fundamental problem with heteroscedastic errors. Fig. 6 displays the residual vs. prediction error plot, though no clear pattern is visible, however, the Jarque-Bera (JB) normality assumption test was performed to ensure the correctness of our assumption. According to Fig. 6, the pandemic caused an early decline in prices throughout 2020–21, followed by a steep rise as producers reduced supply and demand soared. The assumption is satisfied because the Durbin-Watson (DW) test result of 1.98 indicates that there is no autocorrelation. However, the Breusch-Godfrey (BG) Test was employed too, which identifies the autocorrelation up to any predetermined order p. The null hypothesis ($H_0$) of BG shows no serial correlation of any order up to p.

TABLE 6
BREUSCH-GODFREY TEST FOR AUTOCORRELATION

| Statistic | p-value | f-value | f p-value | lag |
|---|---|---|---|---|
| 36.52 | 0.00 | 1.124 | 0.362 | 6 |
| 39.48 | 0.00 | 0.926 | 0.530 | 12 |
| 52.48 | 0.00 | 1.806 | 0.064 | 24 |
| 65.68 | 0.00 | 11.104 | 0.006 | 48 |
| 65.90 | 0.065 | 22.323 | 0.012 | 50 |

Table 6 displays the test statistic $x^2 = 36.52$ and $p-value = 0.000$, indicating we can reject $H_0$ and conclude that autocorrelation exists among the residuals at some order less than or equal to 6 lags. We have tested 12, 24, 48, and 50 lags and found a $p-value > 0.05$ at lag 50, where $H_0$ cannot be rejected. However, considering the seasonal correlation, we have considered adding seasonal dummy variables to the model.

Following that, a normality test was run on the residuals, with the premise that the model's residuals are normally distributed. Observing the histogram plot (Fig. 4), we observe that the distribution of the residuals roughly resembles a bell shape, although there are a few large outliers that could lead to a significant skewness.

According to Fig. 4, the distribution of the residuals is bell-shaped. However, to ensure the normality assumption, we checked the QQ plot displayed in Fig. 5 followed by statistical tests (Table 7). QQ plot indicates a non-normal residual distribution.

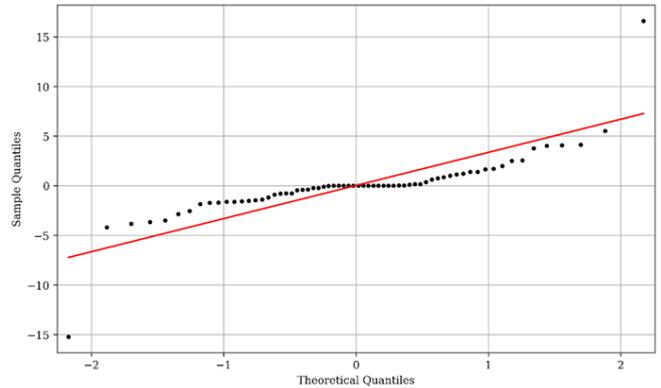

Fig. 5. QQ plot of residuals

TABLE 7
NORMALITY TEST

| Description | Statistics | p-value | output |
|---|---|---|---|
| Shapiro-Wilk Test | 0.727 | 0.00*** | data does not look normal (reject H0) |
| D'Agostino's K2 Test | 30.99 | 0.000*** | data does not look normal (reject H0) |
| Jarque Bera test (JB) | 485.864 | 0.000*** | data does not look normal (reject H0) |
| Anderson Darling test | 4.764 | | critical values = array ([0.546, 0.622, 0.746, 0.870, 1.035]); significance level=array ([15., 10., 5., 2, 1.]). The test results are significant at every significant level, which means H0 can be rejected. Thus, data is not normally distributed. |

Fig. 6 displays the regression residuals and fitted series. Numerous significant outliers can be seen in the graph, but the largest one is in 2020.

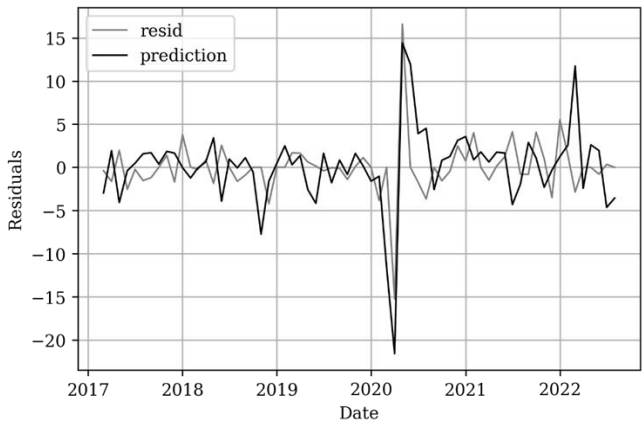

Fig. 6: Regression Residuals and Fitted Series

Table 8 displays the values for the residuals studied to determine the precise dates when the largest outliers were realized. It is evident that the two most extreme residuals were in April'20 (-15.24) and May'20 (16.59). These residuals represent unique or critical events, outliers, or anomalies in the data that have a big impact on WTI returns. The inclusion of dummy variables for these residuals allows the model to adjust and account for these influential observations properly.

TABLE 8
DUMMY VARIABLES CONSTRUCTION

| Date | Smallest residuals |
|---|---|
| Dummy exogeneous variables | |
| 2020-04-01 | -15.243 |
| 2020-05-01 | 16.589 |

Due to the perfect fit of the dummy variables to the two extremely outlying observations, the rerun of the regression along with the dummy variables significantly increased the pseudo $R^2$ value from 0.58 to 0.72. Appendix II reports the estimates of the QR. The distributions were divided into four different quantiles (i.e., $\tau = 0.25, 0.50, 0.75, \& 0.90$) to get a mixed variety of low, medium, and high return conditions. Fig. 7 displays the diagnostic plot, where it can be observed that the errors follow a normal distribution. This has effectively established a baseline model to estimate the effect of the event on our target variable.

Furthermore, both missing variables and an inappropriate functional form were discovered using the RESET (Ramsey Regression Equation Specification Error Test). An F-value of 0.248 and a corresponding p-value of 0.620 from the data show that we cannot rule out $H_0$ that the model contains no omitted variables. To ascertain whether there is a structural break in the data at any given moment, the CUSUM test (Ploberger & Krämer, 1992) for parameter stability based on OLS residuals was carried out.

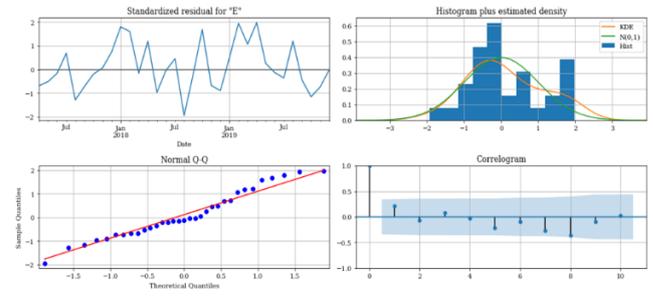

Fig. 7. Residuals diagnostics

Table 10 presents the cumulative total and cumulative sum of squares of recursive residuals to test the structural stability of the models. The absence of any structural breaks is the null hypothesis. The test statistic and associated p-value (0.90) suggest that $H_0$ cannot be rejected, and the coefficients are stable over time; this confirms that the model does not have a structural break for any possible break date in the sample.

TABLE 9
PARAMETER STABILITY TEST

| test statistic | 0.56 |
|---|---|
| p-value | 0.905 |
| Critical values | [(1, 1.63), (5, 1.36), (10, 1.22)])] |

A. *Causality analysis:*

Causal Impact Analysis reduces the noise and provides real statistical insight which leads to the confidence to move forward with. The average value of the response variable is 1.36. If the intervention had not occurred, it was expected that the average response would have been 3. 21. The response variable had an overall value of 43.6 when the post-intervention period's individual data points were added together. But if the intervention had not happened, we would have anticipated a total of 116.77 in absolute terms, with a confidence interval of [80.29, 154.44].

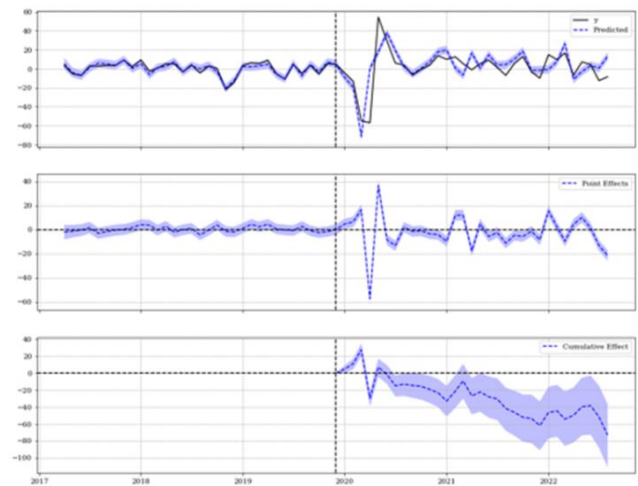

Fig. 8: Causal Impact plot

With an upper and lower bound of [-94.96, -31.46], the response variable showed a relative decline of - 62.7%. This demonstrates that the detrimental impact seen during the intervention period is statistically significant. Fig. 8 displays the causal impact analysis plot. The Bayesian one-sided tail-area probability of getting this result by chance is exceedingly low ($p = 0.0$). This indicates that the causal effect is statistically significant.

## 6. EMPIRICAL RESULTS & DISCUSSIONS

The quantile analysis found the following intriguing trends: The fact that PROD, INFLATION, GPE, and GEPU have a positive and significant impact on the ROC at both the 25% and 50% levels suggests that the relationship is robust and not just limited to a particular quantile level. This implies that when the market is bullish, these variables have a substantial impact on the return on the asset, and investors need to take these factors into consideration when making investment decisions. The intercept term appeared negative for the lower and median quantiles, which suggests that, on average, WTI returns are negative or below zero at these quantiles, even when the predictor variables are set to zero. This is primarily because of pandemic panic and supply chain disruption during the pandemic phase. Table 10 presents a complete discussion on each factor based on the QR analysis displayed in Table 9.

TABLE 10
EMPIRICAL ESTIMATION

| Variables | Causality analysis |
|---|---|
| SP | The negative estimate of the coefficient implies that at the $50^{th}$ quantile, the SP return has a negative effect on the WTI return, whereas at other quantiles, there is no meaningful effect. This seems logical considering the specific combination of conditions that led to this link between the SP return and the WTI return at the $50^{th}$ quantile during and after the pandemic. Wang et al., (2020) discovered a statistically significant positive connection at lower quantiles in their work. But Dutta & Raunak (2020) found a strong negative relationship between SP and WTI returns during the pandemic, which is in line with our findings. |
| PROD | The coefficient estimates for INDPRO are significant across all quantiles of the WTI return distribution, suggesting the relationship with WTI returns is consistent across different levels of returns. This implies that a strong industrial sector is associated with a higher ROC. This is in line with Kalymbetova (2021), and Ratti & Vespignani (2016), who found a positive cointegrated relationship between the INDPRO and oil prices. |
| CURRENCY | Positive and statistically significant estimates at the median and 3rd quantile show that when the value of the US dollar goes up, WTI returns go up at the median and 3rd quantile of the distribution. Similar findings were reported by Olayeni et al. (2020), and Singhal et. al. (2019). The median and 3rd quantiles of our data set correspond to the height of a devastating pandemic supply chain disruption. One possible inference from this relationship is that changes in the value of the US dollar can affect the price of WTI, which in turn can have implications for the wider economy. |
| M1SL | The effect of M1SL on WTI's return is strongest at the median level but weaker at other levels. This finding may have important implications for investment strategies. The investors may want to adjust their investment strategies, accordingly, depending on whether they expect WTI return to be below or above the median level. However, no evidence of a long-run relationship can be drawn from this. This finding supports a recently concluded study from Sørensen & Johansen (2021), who applied cointegration tests on US assets and the money supply. |
| UNRATE | Several studies (e.g., Hammoudeh & Li 2012; Li & Li 2019; Nguyen & Sriananthakumar 2019) have looked at the link between the unemployment rate and the WTI return, and they have come to different conclusions about the size and direction of the link. During the time span of our investigation, we found no statistically significant effect. However, additional research is required to completely comprehend the nature of this link and its operating processes, which is outside the scope of this work. |
| INFLATION | There is a strong and positive link between inflation and WTI return in the 1st and middle quantiles. Even though Kilian (2014) presented a complete analysis of the elements that contribute to oil price volatility, including inflation, and implied that the link between these variables can be influenced by a variety of supply and demand factors in the oil market, Wang et al., (2019) evaluated the relationship between the oil price and inflation and reported a positive and statistically significant relationship between both variables at specific quantiles. |
| WUPI | The epidemic had no meaningful effect on the WTI return across all quantiles of our data. This suggests that, while the pandemic caused some volatility in the WTI price, it did not produce a continuous trend in either direction that would have had a significant impact on the WTI return. Our findings are consistent with those of Liu et al., 2020; Narayan (2020); and Zhang & Hamori, S. (2021), who found that the pandemic had no effect on WTI returns across all quantiles. |
| GPE | All the quantiles show favorable and significant outcomes of GPE. This demonstrates the relationship between the global energy price index and the WTI return, both of which are impacted by the dynamics of supply and demand, geopolitical events, and global economic conditions. Given the close relationship between the WTI return and the global price of energy index, this positive relationship is not unexpected |
| GEPU | The study reveals that the influence of GEPU on WTI return is stronger in the lower and middle ranges of the WTI return distribution, but weaker in the upper range. This suggests that economic policy uncertainty can significantly affect the oil market during market volatility or stress, while its impact may be less pronounced during market stability or good performance. This is consistent with the findings of Li & Yang 2020. |
| VIX | VIX indicates a significant negative correlation between WTI return and volatility, indicating an inverse relationship between volatility and returns in financial markets. Increased volatility leads to risk-averse investors selling off riskier assets, potentially resulting in lower returns. |
| GPR | The study reveals that geopolitical risk's impact on WTI return may be stronger at certain levels of the WTI return distribution. This finding is consistent with the idea that GPR can have a more pronounced impact on the oil market during times of market uncertainty or instability, when investors are more sensitive to political and economic events. At the same time, the impact of GPR may be less pronounced during periods of market stability or when the market is performing well. This finding is consistent with Wang et. al. 2017; Ye & Zyren 2015. |
| SPREAD | SPREAD is significant and positive in the 50th and 90th quantiles. This is like and in line with the study by Bampinas & Panagiotidis (2019), who specifically examined the relationship between oil and stock market returns through QR and reported similar findings. It suggests that SPREAD yields have a stronger effect on WTI returns when the returns are in the upper half or top decile of their distribution. |

The critical findings are summarized as:
- the market return (erSP) has a negative effect on crude oil returns at the median and 90th quantiles, but not at the lower or higher quantiles. Production (dPROD), global economic policy uncertainty (dGPE), and the treasury yield curve (dSPREAD) all have positive effects on crude oil returns across all quantiles.
- the money supply (dMONEY) has a large negative effect on crude oil returns at the 25th quantile.
- dUNRATRE has a positive effect on WTI returns, but not significant at any of the quantiles (Qn 0.25, Qn 0.5, Qn 0.75, and Qn 0.9). This indicates that the unemployment rate may have some influence on WTI returns but does not reach statistical significance in this model.
- dINFLATION has a considerable positive effect on crude oil returns at the 25th and 50th quantiles, but not at higher quantiles. It is possible that the correlation between the inflation rate and WTI returns is nonlinear. The inflation rate may have a greater effect on returns while they are lower (e.g., during recessions), but as returns rise (e.g., during expansions of the economy), its effect may become less pronounced or level out.
- the VIX volatility index (dVIX) has a major negative impact on crude oil returns at the 25th, 50th and 90th quantile. Crude oil returns typically suffer negative effects at various levels of the WTI return distribution when market volatility and fear are high (as evidenced by a higher VIX).
- other factors, such as currency exchange rates (dCURRENCY), the pandemic index (dWUPI), and the geo-political risk (dGPR), have mixed or minor effects on crude oil returns across quantiles.
- the returns on crude oil at all quantiles are significantly impacted by the month dummies D_Apr20 and D_May20. The significance of variables suggests that these extreme deviations have a significant effect on the overall relationship between the predictors and WTI returns.

The pseudo R2 values are high, indicating that the model fits the data well.

Since economic theory does not say which parts or how many should be used in the study, there are many possible variables that could be considered. Our empirical findings have implications for portfolio design and risk management for investors. It also has significant implications for risk management decisions involving hedging and downside risk, given that the financial utility of oil varies depending on market conditions. Finally, our findings have implications for the forecasting of COP across quantiles based on macroeconomic and financial variables. Furthermore, changes in the several parameters considered for this study account for almost 2/3 of the monthly fluctuation in the excess returns

## 7. CONCLUSION

The study used an asset pricing model that combined Arbitrage Pricing Theory (APT) and Quantile Regression (QR) to assess the risk-return relationship of WTI crude oil. To evaluate the risk-return connection of WTI crude oil, the model used multivariate risk components and market returns (SP 500). The report finds that market return, industrial production, global economic policy uncertainty, and the Treasury yield curve have significant positive effects on crude oil returns across all quantiles. The study reveals that the money supply, unemployment rate, inflation rate, and VIX volatility index have significant negative and positive effects on the returns of WTI at different quantiles. The combination of APT and QR provides a comprehensive understanding of the risk-return relationship of the WTI, capturing both linear and nonlinear relationships. The study found that the SP 500 market return is not a significant predictor of WTI returns, suggesting a weak or non-linear relationship. Other key factors, such as PROD, inflation, GPE, and GEPU, have a more significant impact on WTI returns. However, the analysis's time horizon may be too short to detect a significant relationship, as the relationship between the SP 500 return and the WTI return is influenced by longer-term economic or geopolitical factors. The results can help identify profitable investment opportunities and make strategic investment decisions. However, building a trustworthy empirical model requires iteration and is not a precise science.

## REFERENCES

[1] Adekoya, O. B., Oliyide, J. A., Kenku, O. T., & Al-Faryan, M. A. S. (2022). Comparative response of global energy firm stocks to uncertainties from the crude oil market, stock market, and economic policy. Resources Policy, 79, 103004.
[2] Aloui, C., & Hammoudeh, S. (2015). A time-varying copula approach to oil and stock market dependence: The case of transition economies. Energy Economics, 50, 106-118.
[3] Aloui, D., Goutte, S., Guesmi, K., & Hchaichi, R. (2020). COVID 19's impact on crude oil and natural gas S&P GS Indexes.
[4] Alqahtani, A., Bouri, E., & Vo, X. V. (2020). Predictability of GCC stock returns: The role of geopolitical risk and crude oil returns. Economic Analysis and Policy, 68, 239-249.
[5] Alqaralleh, H. (2020). Stock return-inflation nexus; revisited evidence based on nonlinear ARDL. Journal of Applied Economics, 23(1), 66-74.
[6] Bampinas, G., & Panagiotidis, T. (2019). Oil and stock market returns: A quantile regression analysis. International Review of Financial Analysis, 66, 101359.
[7] Beckmann, J., & Czudaj, R. (2013). Oil prices and effective dollar exchange rates. International Review of Economics & Finance, 27, 621-636.

[8] Chan, Y. T., & Dong, Y. (2022). How does oil price volatility affect unemployment rates? A dynamic stochastic general equilibrium model. Economic Modelling, 114, 105935.

[9] Chen, X. (2022). Are the shocks of EPU, VIX, and GPR indexes on the oil-stock nexus alike? A time-frequency analysis. Applied Economics, 1-16.

[10] Chua, R. Y., & Yang, H. (2021). COVID-19 and oil price: Evidence from wavelet coherence analysis. Energy Economics, 105, 105062.

[11] Dai, Z., & Kang, J. (2021). Bond yield and crude oil prices predictability. Energy Economics, 97, 105205.

[12] Dutta, A., & Raunak, R. (2020). Effect of COVID-19 pandemic on stock market and crude oil prices. International Journal of Energy Economics and Policy, 10(4), 431-436.

[13] Ferrer, R., Shahzad, S. J. H., López, R., & Jareño, F. (2018). Time and frequency dynamics of connectedness between renewable energy stocks and crude oil prices. Energy Economics, 76, 1-20.

[14] Fung, W., & Hsieh, D. A. (2004). Hedge fund benchmarks: A risk-based approach. Financial Analysts Journal, 60(5), 65-80.

[15] Guo, Y., Yu, C., Zhang, H., & Cheng, H. (2021). Asymmetric between oil prices and renewable energy consumption in the G7 countries. Energy, 226, 120319.

[16] Hamdi, B., Aloui, M., Alqahtani, F., & Tiwari, A. (2019). Relationship between the oil price volatility and sectoral stock markets in oil-exporting economies: Evidence from wavelet nonlinear denoised based quantile and Granger-causality analysis. Energy Economics, 80, 536-552.

[17] Hammoudeh, S. M., & Li, H. (2012). Sudden changes in volatility in crude oil markets: the impact of the 2008 global financial crisis. Energy Economics, 34(6), 2015-2021.

[18] Herrera, A. M., Karaki, M. B., & Rangaraju, S. K. (2019). Oil price shocks and US economic activity. Energy policy, 129, 89-99.

[19] Husaini, D. H., & Lean, H. H. (2021). Asymmetric impact of oil price and exchange rate on disaggregation price inflation. Resources Policy, 73, 102175.

[20] Jurek, J. W., & Stafford, E. (2015). The cost of capital for alternative investments. The Journal of Finance, 70(5), 2185-2226.

[21] Kalymbetova, A. (2021). The effect of oil prices on industrial production in oil-importing countries: panel cointegration test. International Journal of Energy Economics and Policy.

[22] Kilian, L. (2014). Oil price shocks: Causes and consequences. Annu. Rev. Resour. Econ., 6(1), 133-154.

[23] Köse, N., & Ünal, E. (2021). The effects of the oil price and oil price volatility on inflation in Turkey. Energy, 226, 120392.

[24] Le, T.H. and Chang, Y. (2013), "Oil price shocks and trade imbalances", Energy Economics, Vol. 36, pp. 78-96.

[25] Li, J., & Li, B. (2019). Does the US unemployment rate affect crude oil prices? Evidence from a quantile-on-quantile regression approach. Energy Economics, 84, 104511.

[26] Liu, L., Wang, E. Z., & Lee, C. C. (2020). Impact of the COVID-19 pandemic on the crude oil and stock markets in the US: A time-varying analysis. Energy Research Letters, 1(1), 13154.

[27] Mahmoudi, M., & Ghaneei, H. (2022). Detection of structural regimes and analyzing the impact of crude oil market on Canadian stock market: Markov regime-switching approach. Studies in Economics and Finance

[28] Mahmoudi, R., & Ghaneei, A. (2022). The dynamics of crude oil prices and the role of economic, financial, and geopolitical factors: A Markov switching approach. Energy Economics, 104, 106956.

[29] Malik, F., & Umar, Z. (2019). Dynamic connectedness of oil price shocks and exchange rates. Energy Economics, 84, 104501.

[30] McMillan, D. G., Ziadat, S. A., & Herbst, P. (2021). The role of oil as a determinant of stock market interdependence: the case of the USA and GCC. Energy Economics, 95, 105102.

[31] Mokni, K. (2020). A dynamic quantile regression model for the relationship between oil price and stock markets in oil-importing and oil-exporting countries. Energy, 213, 118639.

[32] Narayan, P. K. (2020). Oil price news and COVID-19—Is there any connection?. Energy Research Letters, 1(1).

[33] Nayar, J. (2020). Integration of oil with macroeconomic indicators and policy challenges in regard to Oman. International Journal of Energy Sector Management, 14(1), 172-192.

[34] Nguyen, D. K., & Sriananthakumar, S. (2019). The effect of money supply on crude oil prices: Evidence from the Bayesian quantile regression approach. Energy Economics, 83, 375-388.

[35] Olayeni, O. R., Tiwari, A. K., & Wohar, M. E. (2020). Global economic activity, crude oil price and production, stock market behaviour and the Nigeria-US exchange rate. Energy economics, 92, 104938.

[36] Pan, Z., Wang, Y., Wu, C. and Yin, L. (2017), "Oil price volatility and macroeconomic fundamentals: a regime switching GARCH-MIDAS model", Journal of Empirical Finance, Vol. 43, pp. 130-142.

[37] Ploberger, W., & Krämer, W. (1992). The CUSUM test with OLS residuals. Econometrica: Journal of the Econometric Society, 271-285.

[38] Prabheesh, K. P., Padhan, R., & Garg, B. (2020). COVID-19 and the oil price–stock market nexus: Evidence from net oil-importing countries. Energy Research Letters, 1(2).

[39] Ratti, R. A., & Vespignani, J. L. (2016). Oil prices and global factor macroeconomic variables. Energy Economics, 59, 198-212.

[40] Salisu, A. A., Olaniran, A., & Tchankam, J. P. (2022). Oil tail risk and the tail risk of the US Dollar exchange rates. Energy Economics, 109, 105960.

[41] Salisu, A. A., Swaray, R., & Oloko, T. F. (2019). Improving the predictability of the oil–US stock nexus: The role of macroeconomic variables. Economic Modelling, 76, 153-171.

[42] Shahzad, S. J. H., Naeem, M. A., Peng, Z., & Bouri, E. (2021). Asymmetric volatility spillover among Chinese sectors during COVID-19. International Review of Financial Analysis, 75, 101754.

[43] Shahzad, S. J. H., Shahbaz, M., Mahalik, M. K., & Hammoudeh, S. (2019). Modelling oil volatility–uncertainty nexus using realised volatility index. Energy Economics, 81, 380-389.

[44] Shahzad, S.J.H., Raza, S.A., Bhatti, M.I., & Ali, S. (2021). What drives crude oil prices? Evidence from frequency-domain causality and network analysis. Energy Economics, 94, 105118.

[45] Singhal, S., Choudhary, S., & Biswal, P. C. (2019). Return and volatility linkages among International crude oil price, gold price, exchange rate and stock markets: Evidence from Mexico. Resources Policy, 60, 255-261.

[46] Sørensen, J., & Johansen, V. (2021). Relationship Between Money Supply and Asset Prices in Developed Countries (Master's thesis, uis).

[47] Wang, K. H., Liu, L., Li, X., & Oana-Ramona, L. (2022). Do oil price shocks drive unemployment? Evidence from Russia and Canada. Energy, 253, 124107.

[48] Wang, Y., Wu, C., & Yang, L. (2017). The impact of geopolitical risks on crude oil prices: An empirical analysis based on a mixed-frequency model. Energy Economics, 65, 88-98.

[49] Wang, Y., Wu, C., Yang, L., & Zhang, Z. (2019). The nexus between oil price and inflation: New evidence from wavelet-based quantile-in-quantile regression. Energy Economics, 78, 586-603.

[50] Wang, Y., Zhang, X., & Liu, L. (2020). The asymmetric dynamic relationship between crude oil and US stock markets: Evidence from threshold quantile regression analysis. Energy Economics, 92, 104874.

[51] Wei, Y., Qin, S., Li, X., Zhu, S., & Wei, G. (2019). Oil price fluctuation, stock market and macroeconomic fundamentals: Evidence from China before and after the financial crisis. Finance Research Letters, 30, 23-29.

[52] Xiao, Z., Guo, H., and Lam, M. S. (2015). "Quantile Regression and Value at Risk." In Handbook of Financial Econometrics and Statistics, edited by C.-F. Lee and J. Lee, 1143–1167. New York: Springer.

[53] Zhang, D., Ma, F., & Wei, Y. M. (2020). Oil price volatility and stock market returns in China: Evidence from a quantile regression approach. Energy Economics, 91, 102249.

[54] Zhang, W., & Hamori, S. (2021). Crude oil market and stock markets during the COVID-19 pandemic: Evidence from the US, Japan, and Germany. International Review of Financial Analysis, 74, 101702.

[55] Zhao, L. T., Xing, Y. Y., Zhao, Q. R., & Chen, X. H. (2023). Dynamic impacts of online investor sentiment on international crude oil prices. Resources Policy, 82, 103506.

[56] Hair, J.F., Hult, G.T.M., Ringle, C.M. and Sarstedt, M. (2017), A Primer on Partial Least Squares Structural Equation Modeling (PLS-SEM), Sage, Thousand Oaks, CA

[57] Bohdalová, M., & Greguš, M. (2017, September). Impact of uncertainty on European market indices quantile regression approach. In CBU International Conference Proceedings (Vol. 5, pp. 57-61).

# APPENDIX -I

Descriptive statistics of the transformed variables

| Variable | Mean | Median | Max | Min | Std Dev | Skew | Kurtosis | JB | ADF | VIF |
|---|---|---|---|---|---|---|---|---|---|---|
| WTI | 0.75 | 3.38 | 54.55 | -56.82 | 14.59 | -0.98 | 8.10 | 161.13** | -6.76** | 2.60 |
| SPY | 0.90 | 1.58 | 6.29 | -21.04 | 3.95 | -2.88 | 14.00 | 543.34** | -7.07** | 2.29 |
| PROD | 0.01 | -0.03 | 15.75 | -10.31 | 2.76 | 2.16 | 18.51 | 848.54** | -6.36** | 1.95 |
| CURRENCY | 0.00 | 0.00 | 0.03 | -3.21 | 2.01 | -0.11 | 0.49 | 6.49* | -5.46** | 2.14 |
| MONEY | -0.01 | 0.02 | 110.79 | -120.24 | 20.32 | -0.71 | 32.40 | 2466.58** | -6.35** | 2.10 |
| UNRATE | -0.01 | -0.01 | 97.74 | -131.38 | 21.29 | -2.16 | 28.83 | 1955.82** | -6.52** | 2.61 |
| INFLATION | -0.00 | 0.01 | 0.67 | -0.11 | 0.34 | -0.70 | 2.69 | 22.58** | -5.73** | 2.61 |
| PANDEMIC | 0.00 | 0.00 | 4.45 | -2.89 | 0.85 | 1.75 | 14.19 | 501.14** | -7.93** | 1.59 |
| GPE | 0.16 | 1.09 | 46.32 | -0.33 | 11.41 | 0.26 | 4.31 | 42.66** | -7.61** | 3.59 |
| VIX | 0.03 | -0.01 | 2.25 | -0.46 | 0.32 | 4.90 | 32.69 | 2757.41** | -6.77** | 1.29 |
| GPR | 0.06 | 2.41 | 91.15 | -130.65 | 36.32 | -0.41 | 1.48 | 6.34* | -4.74** | 1.83 |
| SPREAD | -0.02 | -0.00 | 0.51 | -0.98 | 0.19 | -1.56 | 9.22 | 222.02** | -5.21** | 1.18 |
| GEPU | 0.83 | -3.34 | 140.48 | -101.33 | 43.40 | 0.63 | 1.48 | 8.78* | -6.10** | 2.14 |

# APPENDIX -II

Parameter Estimates of Risk Factors

| Variable | Qn (0.25) | Qn (0.5) | Qn (0.75) | Qn (0.9) |
|---|---|---|---|---|
| erSP | -0.10 (0.06) | -0.19 **(0.04) | 0.05(0.09) | -0.11(0.19) |
| dPROD | 0.18**(0.06) | 0.26 **(0.04) | 0.16*(0.07) | 0.37**(0.14) |
| dCURRENCY | 0.05(0.03) | 0.06**(0.02) | 0.12**(0.04) | 0.11(0.07) |
| dMONEY | -0.09 (0.04) | -0.08*(0.03) | 0.00(0.07) | 0.14(0.13) |
| dUNRATE | 0.10(0.07) | 0.07 (0.04) | 0.07(0.08) | 0.05(0.16) |
| dINFLATION | 0.17 **(0.06) | 0.10 **(0.04) | 0.14(0.09) | -0.00(0.19) |
| dWUPI | 0.01(0.03) | -0.00(0.02) | 0.06(0.05) | -0.05(0.04) |
| dGPE | 0.29**(0.07) | 0.40**(0.03) | 0.39**(0.05) | 0.38**(0.09) |
| dVIX | -0.25 **(0.11) | -0.30**(0.06) | -0.16(0.11) | -0.53*(0.26) |
| dGPR | 0.00(0.05) | 0.06**(0.03) | 0.06(0.05) | 0.09(0.11) |
| dSPREAD | 0.08(0.05) | 0.14**(0.03) | 0.12(0.06) | 0.29*(0.13) |
| dGEPU | 0.11**(0.02) | 0.08**(0.02) | 0.02(0.05) | 0.04(0.12) |
| D_Apr20 | -23.25**(4.57) | -23.38**(2.54) | -27.25**(5.70) | -27.59*(12.55) |
| D_May20 | 23.85**(4.12) | 26.05**(2.87) | 21.63**(5.66) | 14.94(10.85) |
| Intercept | -0.84**(0.95) | -0.02(0.05) | 1.31**(0.32) | 2.50**(0.53) |
| Pseudo R2 | 0.72 | 0.68 | 0.69 | 0.74 |

Cell entries are coefficients, with standard errors in parentheses, * denotes $p < 0.05$; ** denotes $p < 0.01$